\documentclass[12pt,preprint]{aastex}

\slugcomment{Accepted for publication in The Astronomical Journal;}

\shorttitle{Keck/LRIS Spectroscopy of the z = 0.437 DLA absorber in front of 3C196}
\shortauthors{Gharanfoli et al.}

\received{2006 June 30}
\begin{document}


\title{Emission Line Spectroscopy of a Damped Ly$\alpha$ Absorbing
Galaxy at z = 0.437}


\author{Soheila Gharanfoli and Varsha P. Kulkarni}
\affil{Dept. of Physics and Astronomy, University of South Carolina, Columbia, SC 29208}
\author{Mark R. Chun} 
\affil{Institute for Astronomy, University of Hawaii, Hilo, HI, 96720}
\and
\author{Marianne Takamiya}
\affil{Dept. of Physics and Astronomy, University of Hawaii, Hilo, HI 96720}
\email{}



\begin{abstract}
We present Keck/LRIS spectra of a candidate damped Ly$\alpha$ (DLA) galaxy toward the QSO 3C196 (z$_{em}$ = 0.871). The DLA absorption system has a redshift of z$_{DLA}$ = 0.437, and a galaxy at 1.5$''$ from the QSO has been identified in high resolution imaging with WFPC2/HST. We have detected emission lines of [O~II] $\lambda$3727, H$\beta$, [O III] $\lambda$5007, H$\alpha$ and [N II] $\lambda$6584 at the absorption redshift. Based on the emission lines, we have found the redshift of the galaxy to be z$_{em}$ = 0.4376 $\pm$ 0.0006. The emission lines also enabled us to calculate the extinction-corrected luminosities and metallicity indicators using established indices based on line strengths of different emission lines. These indicators suggest that the ISM of the DLA galaxy has a high metallicity comparable to or perhaps twice as much as solar (e.g. 12+ $\log$(O/H) = 8.98 $\pm$ 0.07). Based on the strengths of H$\alpha$ and on the reddening derived from the relative strengths of H$\alpha$ and H$\beta$, the star formation rate is 4.7 $\pm$ 0.8 M$_{\sun}$ yr$^{-1}$. This places the galaxy in the range of gas-rich spiral galaxies.
\end{abstract}



\keywords{quasars: absorption lines; galaxies: abundances; galaxies: ISM; galaxies: formation}
\clearpage


\section{INTRODUCTION}{\label{intro}}

Damped Lyman-alpha (DLA) absorption systems, observed in spectra of quasars (QSOs), provide a powerful tracer of the chemical history of high redshift galaxies (Pettini {\it et al.} 1997; Prochaska \& Wolfe 2002) and represent a unique sample for studying the interstellar medium (ISM) of distant galaxies. These systems contain mostly neutral gas, with neutral hydrogen column densities $\log$ N$_{\rm HI}$ $\geq$ 20.3, and thus provide an important link in our understanding of the star formation history of the Universe. DLAs remain the only class of galaxies at cosmologically significant redshifts with detailed measurements of element abundances. Unfortunately, the DLA host galaxies are faint and located at small projected distances from the QSOs. The background  QSOs are much brighter than the DLA and thus can contaminate the background at the position of the DLA, making spectroscopic studies of the DLA galaxies a challenging task.

Recent studies combining high angular resolution images and high SNR spectra are providing key evidence revealing the nature of DLAs. So far, most imaging studies of high-z and low-z DLAs have failed to detect large luminous disks (e.g. Jimenez {\it et al.} 1999; Kulkarni {\it et al.} 2000, 2001; Chun {\it et al.} 2006 and references therein). The lack of substantial chemical evolution found in studies of element abundances in DLAs (e.g. Kulkarni {\it et al.} 2005; P\'eroux {\it et al.} 2006 and references therein) suggests that the currently known population of DLAs seems to be dominated by metal-poor objects. Thus DLAs may consist of dwarf or low-surface brightness galaxies with modest star formation rate (SFR). Indeed, most emission-line imaging searches for DLAs also indicate low SFR (e.g. Kulkarni {\it et al.} 2006 and references therein). It is possible however that the metal-rich DLAs are systematically underrepresented because of dust obscuration of background QSOs (e.g. Fall \& Pei 1993; Boisse {\it et al.} 1998).

SFRs can be estimated with several different methods. Emission line imaging provides a direct method to determine the SFR in low-z DLAs, although it is subject to the possibility of dust extinction. An alternative method based on CII$^*$ absorption lines has been suggested by \cite{wolf03a}. However this method is less direct, since it depends on several assumptions. Furthermore, the CII$^*$ method is not accessibble for low-z DLAs, since the CII$^*$ absorption lines fall in the UV. SFRs determined from optical emission lines have been extensively studied in nearby galaxies (e.g. Kennicutt 1998; Kormendy \& Kennicutt 2004) and DLAs (Chen {\it et al.} 2005; Kulkarni {\it et al.} 2006 and references therein). The hydrogen recombination lines are directly related to the UV emission of the shortest-lived stars and thus measuring H$\alpha$ in the optical wavelength range provides a good tool to estimate the SFR. Assuming that the shape of the initial mass function is known, the main uncertainty in the measurement of optical emission line luminosities is extinction. However we can estimate the amount of extinction by measuring the Balmer decrement (e.g. Pei 1992; Cardelli {\it et al.} 1989).

So far, SFR estimates for most of the absorbers appear to be small: 63\% of the detections and about 73\% of the limits in the sample studied by \cite{kulk06} show SFR $<$ 5 M$_{\sun}$ yr$^{-1}$. This can be explained by one of the following possible cases: 1) The SFRs in DLAs are in fact low; 2) dust extinction is high preventing the line  emission to escape from the HII region; 3) the star-forming regions are too close to the QSO line-of-sight masking the true SFRs. In order to distinguish the dominant effect, more emission line imaging studies should be undertaken.

We have carried out a spectroscopic study of a spiral galaxy (galaxy \# 4 from Le Brun {\it et al.} 1997) 1.5$''$ away from 3C196 (Q0809+483). This QSO has a 21 cm absorption line (Brown \& Mitchell 1983) as well as a DLA system at z$_{abs}$ = 0.4368 with $\log$ N$_{\rm HI}$ = 20.8 $\pm$ 0.2 (Boiss\'e {\it et al.} 1998). The spiral galaxy with AB(F702)\footnote[1]{AB magnitude will be used throughout the paper. AB(F702) refers to the AB magnitude in the HST/WFPC2 filter F702W ($\bar{\lambda }$ = 6940.1 \AA).}= 19.9 has been reported as a candidate host galaxy for the DLA (Le Brun {\it et al.} 1997). The galaxy was suggested to be responsible for the DLA in the spectroscopic program of \cite{chen05}, based on the detection of H$\beta$ and [O~II] emission lines in the spectra of the galaxy at z = 0.43745 $\approx$ 138 km s$^{-1}$ away from the redshift of the DLA. They also noted a narrow Ca II H and K absorption doublet in the spectrum of the background QSO at 21 km s$^{-1}$ blueshifted from the systematic redshift of the galaxy. No additional emission features had been reported prior to the present work.

HST/WFPC2 images showed that galaxy \# 4 has the morphology of a barred spiral galaxy with B$-$R = 2.5 (Le Brun {\it et al.} 1997; Boiss\'e \& Boulade 1990). Typical B$-$R colors of normal galaxies at z = 0.5 range from 3.2 for ellipticals to 1.2 for irregular galaxies with B$-$R = 2.1 for Sbc galaxies (Fukugita {\it et al.} 1995). Galaxy \# 4 is thus comparable in B$-$R (though possibly little redder than) to a typical Sbc galaxy at z = 0.5. \cite{lebr97} also detected a compact galaxy (galaxy \# 3) 1.1$''$ away from 3C196. This galaxy has B$-$R = 2.3 (Boiss\'e \& Boulade 1990), which is comparable to the typical B$-$R of a Sbc galaxy at z = 0.4 or z = 0.9. We note however that \cite{lebr97} have suggested that galaxy \# 3 may be associated with the QSO. In this paper, we study the spiral galaxy \# 4, and put limits on the line emissions for galaxy \# 3.

The paper is organized as follows. The observations and data reduction are described in \S \ref{obs+red}. The results are presented in \S \ref{res} and discussed in \S \ref{dis}. Throughout this paper we assume $\Omega_{\Lambda}$ = 0.7, $\Omega_{m}$ = 0.3, and h = 0.7.\\


\section{OBSERVATIONS AND DATA REDUCTION}{\label{obs+red}}

\subsection{Observations}
\begin{figure}
\centering
\includegraphics[scale=1.0]{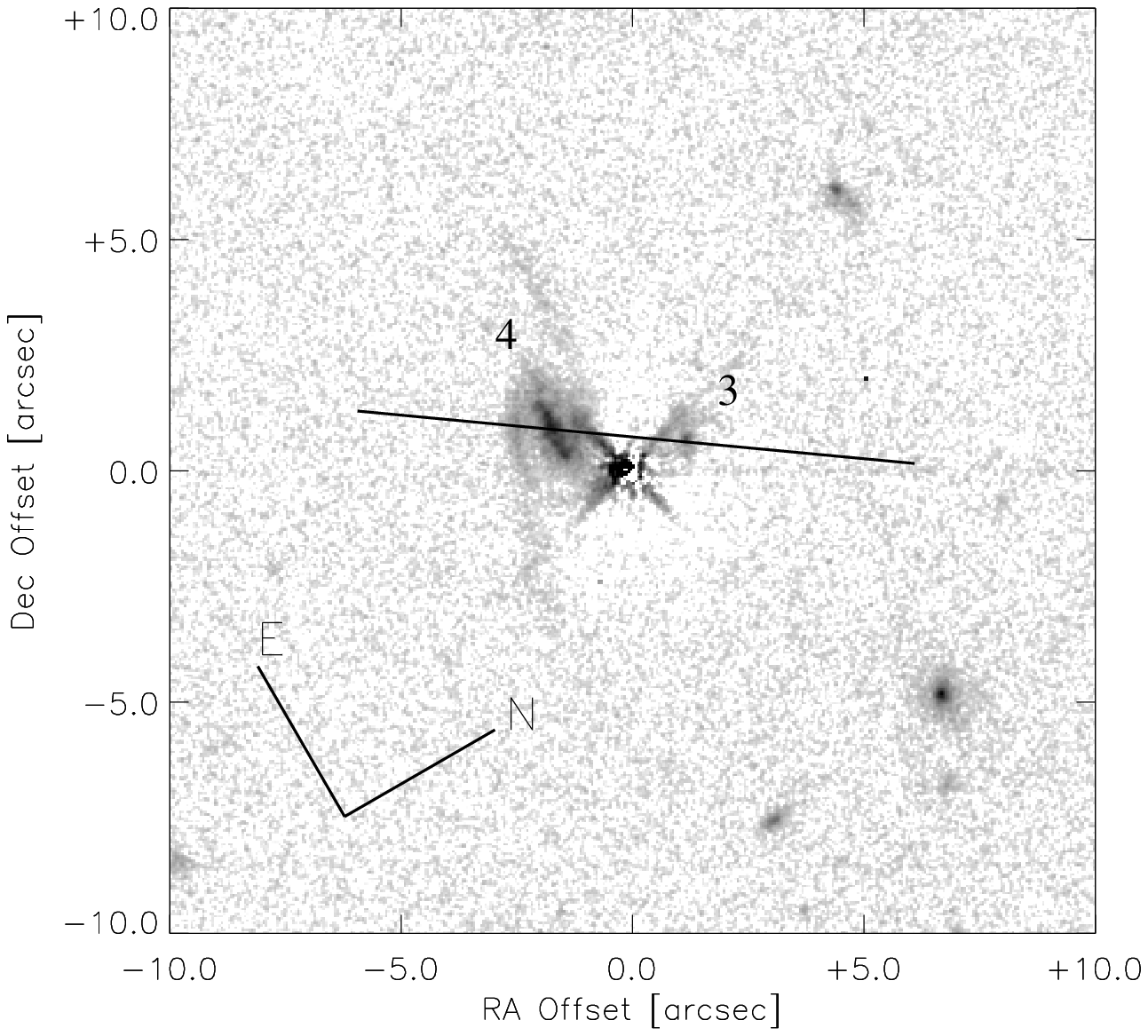}
\caption{\label{fig:f1} WFPC2 image of the field of QSO 3C196 (z$_{em}$ = 0.871) with a DLA at z$_{abs}$ = 0.437 (Le Brun {\it et al.} 1997). The solid line shows the position of the slit during the observations. The slit width is 1.02$''$.}
\end{figure}

{\label{obs}}
The spectra were obtained using the blue and red channels of the Keck Low Resolution Imaging Spectrometer (LRIS; Oke {\it et al.} 1995) on March 7, 2005. Observations were carried out in a series of three 1800 s exposures in each channel. Figure \ref{fig:f1} shows the slit alignment used during the observations. The slit was centered on the QSO and rotated at an angle 36.5$^\circ$ West of North. A short 180 s exposure was obtained at the position of the QSO. This frame was eventually used to remove the QSO contamination from the spectra of the galaxy. Next, the slit was offset by 0.75$''$ perpendicular to the slit length where three 1800 s exposures were obtained. Because the natural seeing ranged between $\sim$ 1$'' - 1.5''$, we used a 1.02$''$ slit width. The slit orientation was chosen to include two DLA candidates from \cite{lebr97} (galaxies \#3 and \#4 in Figure \ref{fig:f1}). The blue channel spectrograph was configured with the 400 lines/mm grism, resulting in a 1.09 \AA \ pix$^{-1}$ dispersion and pixel scale of 0.135$''$. The red channel was configured with the 400 lines/mm grating (1.86 \AA \ pix$^{-1}$, 0.215$''$/pix). The blue and red channel configurations provided a spectral resolution of $\sim$ 10 \AA, as measured for the arc lamp emission lines. The effective wavelength coverage in the blue channel was from 3500 to 5700 \AA, and in the red channel from 6700 to 9700 \AA.

Flat-field frames were obtained in a series of four exposures of 45 s each and five exposures of 20 s each, using an internal quartz lamp for the blue and red channels respectively. Wavelength calibration frames were taken before and after each science exposure using a set of 5 internal arc lamps (Hg, Ar, Ne, Zn, Cd). A spectrophotometric standard star G193-74 (V = 15.70, Oke 1990) was observed for flux calibration in both of the channels.\\

\subsection{Data Reduction}{\label{red}}
Standard procedures were applied to reduce the data using IRAF. Individual object and flat-field frames were first bias-subtracted by using the average of the pre- and post-overscan regions in each frame. A normalized flat-field frame was generated from the average of the flat-field frames. The object frames were then flattened using the normalized flat.

Since the galaxy is an extended faint source lying at small angular separation from the bright PSF of the QSO, the spectra of the galaxy were contaminated by the QSO light. Therefore extracting the galaxy spectra required developing a novel data reduction procedure. To remove the QSO background contribution, the two dimensional spectra centered on the QSO were subtracted from the spectra containing both the Galaxy+QSO light, after suitable scaling and registration. We used the Interactive Data Language (IDL) program Image Display Paradigm-3 (IDP-3; Lytle {\it et al.} 1999) to accomplish this. The procedure is explained in Appendix A. With this method, we were successful in obtaining the background subtracted spectra of galaxy \# 4. We were not able to detect emission from the fainter galaxy \# 3 (M$_B$ = -22.0, Le Brun {\it et al.} 1997).

The QSO, Galaxy+QSO, and the background subtracted galaxy spectra, were then extracted and wavelength calibrated. The extraction was carried out with the task APALL in IRAF\footnote[2]{IRAF is distributed by the National Optical Astronomy Observatories, which are operated by the Association of Universities for Research in Astronomy, Inc., under cooperative agreement with the National Science Foundation.}. The three 1800 s extracted spectra in each channel were stacked to form a combined spectrum after rejecting deviant pixels using an averaged sigma clipping algorithm. The QSO and stacked spectra were then calibrated to an absolute flux scale after determining the sensitivity function from the observed spectrophotometric standard.\\

\section{RESULTS}{\label{res}}

\begin{figure}
\centering
\includegraphics[scale=0.6]{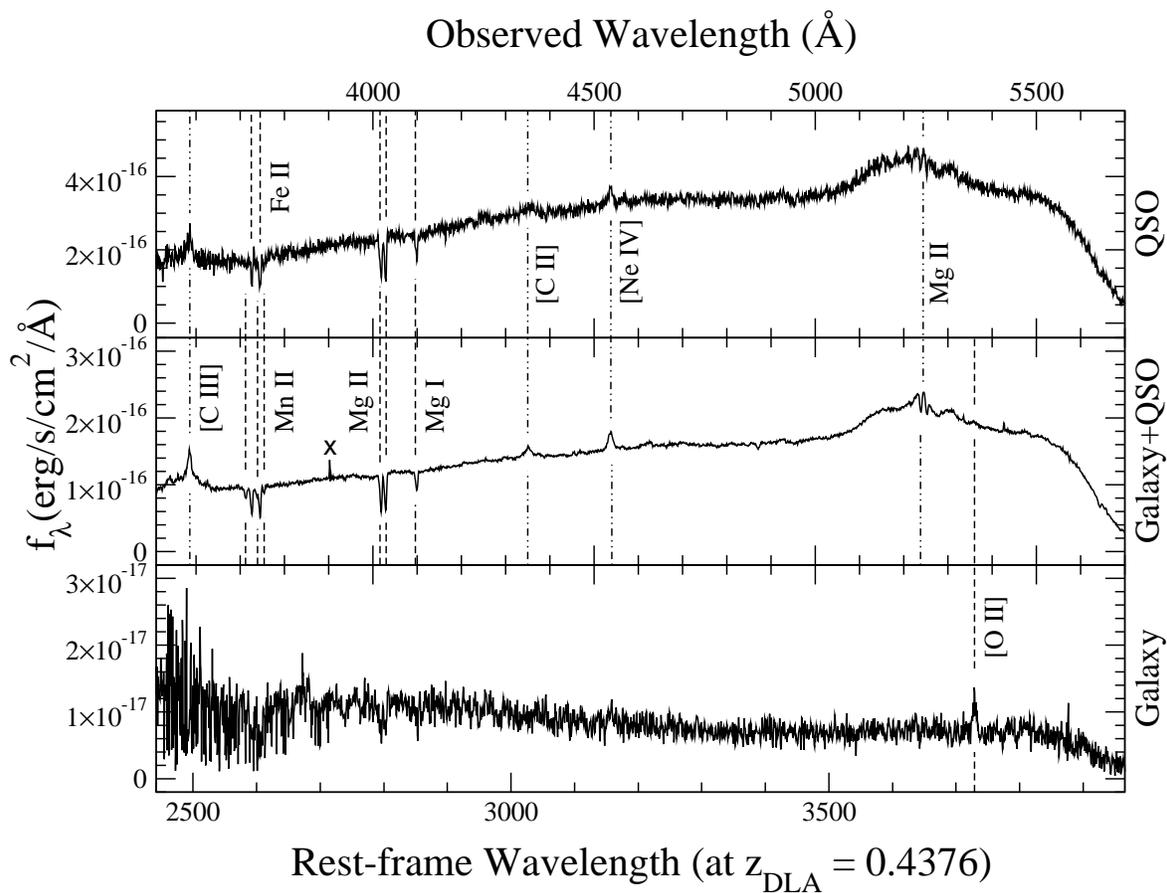}
\caption{\label{fig:f2} The extracted spectra of the QSO 3C196 and its DLA galaxy in the blue channel of LRIS which covers a rest-frame wavelength range from 2400 to 4000 \AA. The panels show the spectrum of the QSO, the contaminated spectrum of the galaxy with the QSO light, and the spectrum of the galaxy after subtracting the QSO contribution. The prominent emission/absorption-line features of the galaxy are shown with dashed lines, and emissions of the QSO with dash-dotted lines. Marked with an 'x' is an emission feature which is likely caused by a cosmic ray.}
\end{figure}

\begin{figure}
\centering
\includegraphics[scale=0.6]{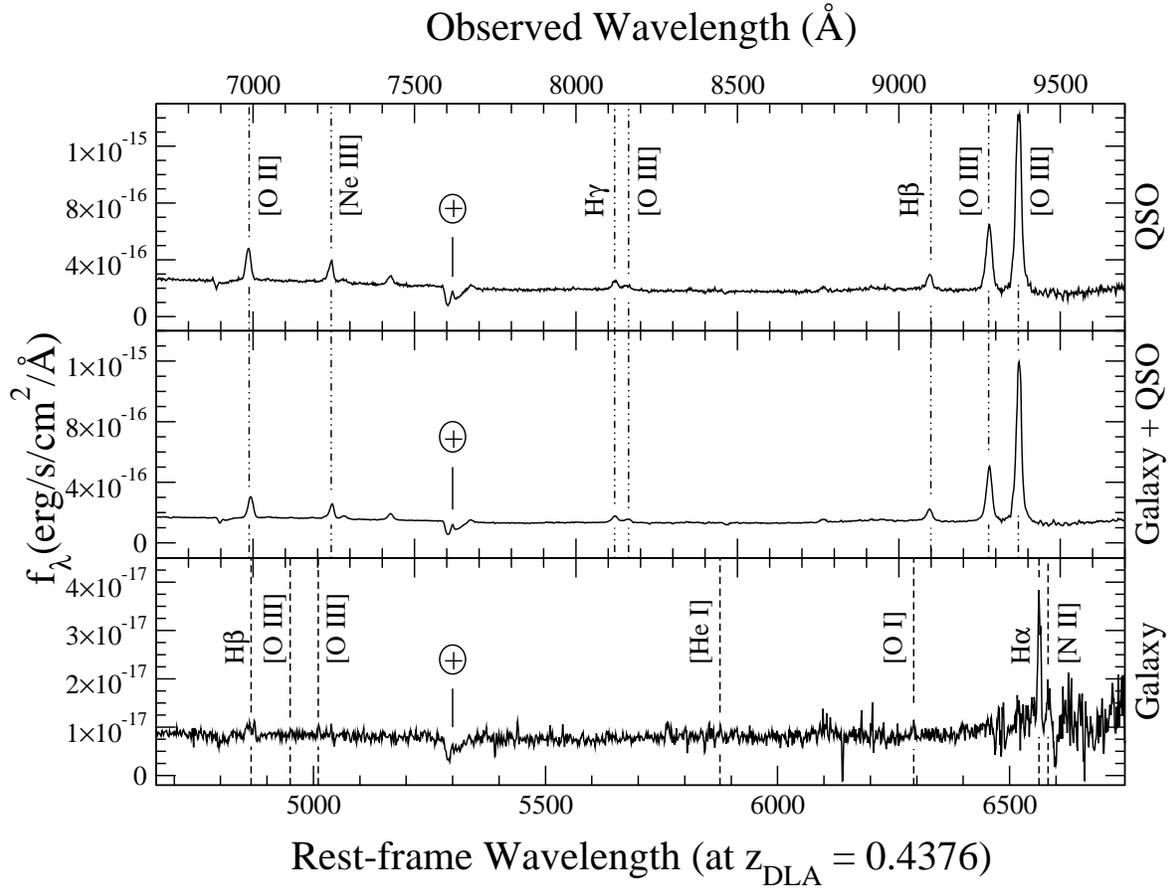}
\caption{\label{fig:f3} The extracted spectra of the QSO 3C196 and its DLA galaxy in the red channel of LRIS which covers a rest-frame wavelength range from 4700 to 6700 \AA. The panels show spectra of the QSO, the galaxy plus the QSO, and the galaxy after subtracting the QSO contribution. The prominent emission/absorption-line features of the galaxy are shown with dashed lines, and emissions of the QSO with dash-dotted lines. The broad feature near 7600 \AA \ is caused by telluric absorption. Note that $H\beta$ emission from the DLA is slightly offset from the [OII] emission of the QSO.}
\end{figure}

We have extracted 3 spectra for each of the channels: the QSO spectrum, the Galaxy+QSO spectrum, and their difference: the Galaxy spectrum. The blue spectra are shown in Figure \ref{fig:f2} and the red, in Figure \ref{fig:f3}. The QSO and Galaxy+QSO spectra are similar which clearly indicates that the spectrum of the galaxy is largely contaminated by the QSO light. In the wavelength calibrated galaxy spectra we identify emission lines at [O II] $\lambda$3727, H$\beta$, [O III] $\lambda \lambda$4959,5007, [He I] $\lambda$5876, [O I] $\lambda$6300, H$\alpha$ and [N II] $\lambda$6584. The H$\alpha$, H$\beta$ and [O II] lines are certain. The others are tentative, since they are comparable to the noise level in the continuum. Identified lines are listed in column 1 of Table \ref{tab:res}. Columns 2 and 3 present the rest-frame and observed wavelengths, respectively, which were then used to calculate the emission redshift (column 4). Due to the presence of uncertainties in the strength of the detected lines, the emission redshift of the galaxy can be calculated by weighting the individual redshifts by the strength of the lines. This gives z$_{em}$ = 0.4376 $\pm$ 0.0006 which is consistent with the DLA absorption redshift in the QSO spectrum. The equivalent widths, observed line-fluxes and calculated luminosities are presented in columns 5, 6 and 7 of Table \ref{tab:res}, respectively.

For galaxy \# 3, assuming the galaxy is at the same redshift as galaxy \# 4, we placed a 3 $\sigma$ upper limit to the H$\alpha$ and [O II] line fluxes of 4.8 $\times 10^{-17}$ ergs s$^{-1}$ cm$^{-2}$ and 1.2 $\times 10^{-17}$ ergs s$^{-1}$ cm$^{-2}$ respectively.

\subsection{Dust Extinction}
Since we have detected both H$\alpha$ and H$\beta$ at the same position of the galaxy, the amount of dust extinction can be readily calculated. We estimated the extinction correction assuming the Small Magellanic Cloud (SMC) type law using the parameterization of \cite{pei92}. This assumption was inspired by a study of more than 800 Mg II absorption systems with 1 $\leq$ z$_{abs} \leq$ 2 in the spectra of Sloan Digital Sky Survey QSOs by \cite{york06}, which showed that the average extinction curve in these systems is similar to the SMC extinction curve with no 2175-\AA \ feature and a rising ultraviolet (UV) extinction below 2200 \AA. Adopting the color excess formula
\begin{equation}
E(B-V) = \frac{1.086}{k(H\beta)-k(H\alpha)} \ln \Biggl( \frac{H\alpha /H\beta}{2.88} \Biggr)
\end{equation}
with the parameters k(H$\beta$) and k(H$\alpha$) taken from \cite{pei92}, we estimated E(B$-$V) = 0.54 $\pm$ 0.08. The corrected luminosities L$^o$ and corresponding uncertainties are presented in column 8 of Table \ref{tab:res}. The uncertainties in L$^o$ were calculated by propagating the errors in luminosities and the extinction factor, which in turn were calculated by using the measurement errors in H$\alpha$ and H$\beta$ fluxes. The uncertainty in the extinction factor is the dominant term. As a comparison, we have also calculated Milky Way (MW) type extinction corrections (Pei {\it et al.} 1992). These corrections are listed in the last column of Table \ref{tab:res} and suggest that MW type extinction corrections are 9$ - 15\%$ higher than SMC type corrections: E(B$-$V) = 0.56 $\pm$ 0.08 for MW type corrections.

It is interesting to compare the above E(B$-$V) with the reddening of the QSO. To do this, we used the color excess $\Delta (g-i) = (g-i)_{obs} - (g-i)_{med}$ where $(g-i)_{med}$ is the median $(g-i)$ color for the SDSS quasar composite at the redshift of the QSO, taken from Table 3 of \cite{rich01}. For the QSO 3C196, we estimated $\Delta (g-i)$ = 0.46.
We then used the relation for the SMC reddening law (Richards, {\it et al.} 2003) E(B$-$V)$_{g-i} = \Delta (g-i)(1+z)^{-1.2}$/1.506 where z is the redshift of the absorber. For the DLA toward QSO 3C196, we obtained E(B$-$V)$_{g-i}$ = 0.20. Thus the gas directly in front of the QSO appears to be less dusty than the DLA's gas probed at the line of sight in our emission line spectroscopy.

In the above flux estimations, we did not apply slit corrections. We have estimated the corrections resulting from losses due to a finite slit width. Assuming a Gaussian profile for the standard star, we estimated that 70\% of the light of the standard star is contained within the slit. For the galaxy, integrating over an exponential profile function, resulted an estimation of 40\% of the light of the galaxy within the slit. Combining these two factors, we estimated a total slit correction factor of $\sim$ 1.8. As a result, the emission line fluxes may have been underestimated, but relative line flux ratios remain unchanged. However, we don't know the surface brightness profile of the galaxy precisely and whether the emission lines have the same spatial distribution as the continuum stellar light, or whether the dust has a uniform distribution in the gas. We therefore regard the slit correction a very rough estimate and prefer not to apply it on the extinction corrected line luminosities.

\subsection{Metallicity}
In order to estimate the metallicity of the galaxy, we determined the R$_{23}$ (Pagel {\it et al.} 1979), N2 and O3N2 (Pettini \& Pagel 2004) indices:
\begin{equation}
\log R_{23}\equiv \log ([O II] \lambda 3727+[O III] \lambda \lambda 4959, 5007)/H\beta = 0.28\pm 0.16,
\end{equation}
\begin{equation}
N2\equiv \log [N II]\lambda 6584/H\alpha = - 0.57\pm 0.14,
\end{equation}
\begin{equation}
O3N2 \equiv \log {([O III]\lambda 5007/H\beta)/([N II]\lambda 6584/H\alpha)} = - 0.037\pm 0.21,
\end{equation}
where the errors represent 1$\sigma$ uncertainties.

To determine to which branch of the R$_{23}$ vs. metallicity relation the galaxy belongs, we measured the [O III]$\lambda$5007/[N II]$\lambda$ 6584 ratio:
\begin{equation}
[O III]\lambda 5007/[N II]\lambda 6584 = 0.32\pm 0.12
\end{equation}
Based on the prescription of \cite{kobu99}, the above ratio suggests that the DLA follows the metal-rich branch of the R$_{23}$ vs. metallicity relation. Using the analytic expression of \cite{mcga91} and \cite{kobu99}, we determined 12 + $\log$(O/H) = 8.98 $\pm$ 0.07 $\pm$ 0.15 where the first error comes from flux measurement errors in our data and the second is the reported error by \cite{kobu99}, caused by the uncertainties in the empirical calibration. As an alternative we calculated the oxygen abundance of the galaxy using the semi-empirical relations based on the N2, O3N2 indices from \cite{pepa04}. We obtained 12 + $\log$(O/H) = 8.56 $\pm$ 0.12 $\pm$ 0.18 for the N2 relation and 12 + $\log$(O/H) = 8.74 $\pm$ 0.07 $\pm$ 0.14 for O3N2, where again the first and second errors denote measurement errors in our data and uncertainties in the calibration respectively. These values show that the ISM of the DLA galaxy has a high metallicity, at least 80$\%$ and at most twice as much as solar (12 + $\log$(O/H)$_{\sun}$ = 8.66; Allende-Prieto {\it et al.} 2001; Asplund {\it et al.} 2004).

Applying the MW type extinction-corrected line luminosities in the above calculations, the results remain unchanged, except for the R$_{23}$ ratio, which we determined to be $\log$ R$_{23}$ = 0.26 $\pm$ 0.17, however this does not affect our metallicity estimations.

\subsection{Star Formation Rate}
We calculated the SFR of the galaxy following the prescription of \cite{kenn98} based on a Salpeter IMF:
\begin{equation}
SFR(M_{\sun} \; yr^{-1})=7.9 \times 10^{-42} L(H\alpha) \; (ergs \; s^{-1})
\end{equation}

For the SMC type extinction correction, we obtained SFR $\approx$ 4.7 $\pm$ 0.8 M$_{\sun}$ yr$^{-1}$, which places the galaxy in the range of gas-rich spiral galaxies (James {\it et al.} 2004). In the case of MW type extinction correction, we calculated SFR $\approx$ 5.4 $\pm$ 1.0 M$_{\sun}$ yr$^{-1}$.\\

\section{DISCUSSION}{\label{dis}}
Table \ref{tab:sum} summarizes the estimated properties of the DLA galaxy. It is instructive to compare these properties with those of other galaxies and DLAs.

\subsection{Comparison with other galaxies}
\begin{figure}
\centering
\includegraphics[scale=0.6]{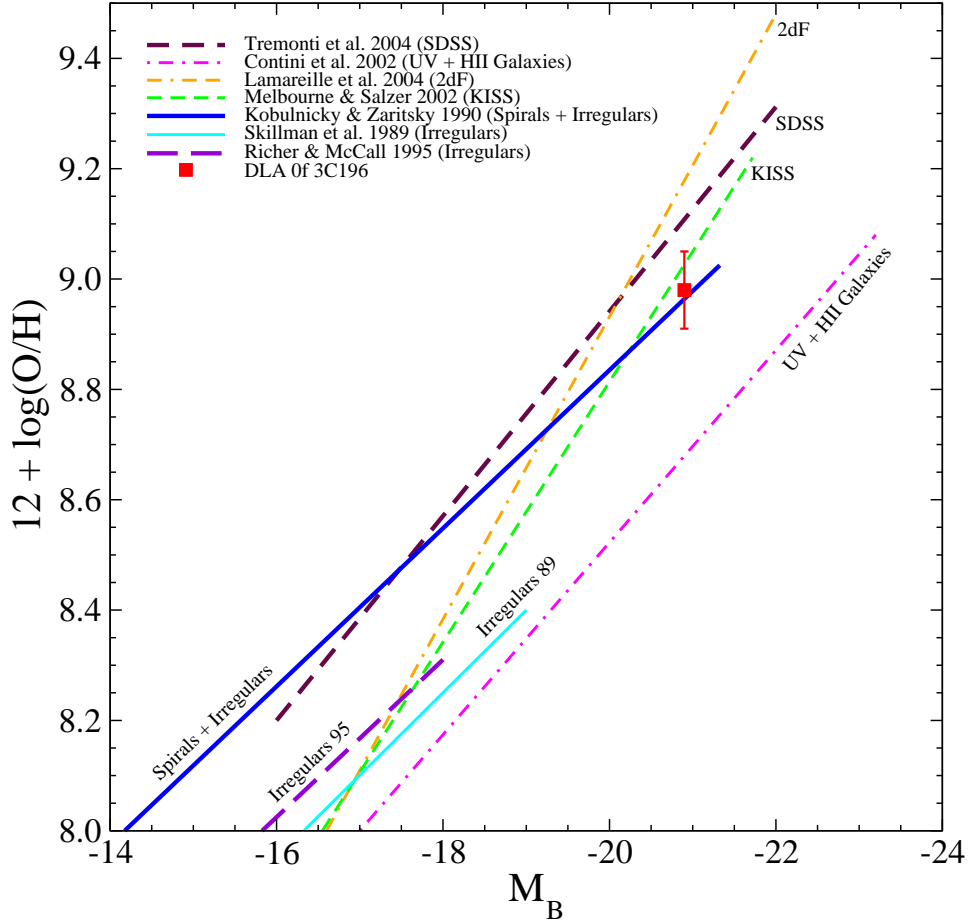}
\caption{\label{fig:f4} Luminosity-metallicity relation for SDSS galaxies and various galaxy samples at z $<$ 0.25, from the compilation by \cite{trem04}. The estimated oxygen abundance 12 + $\log$(O/H) = 8.98 $\pm$ 0.07 (derived from R$_{23}$ index) for the DLA galaxy toward 3C196 together with its luminosity M$_B$(AB) = -21.1 is consistent with the luminosity-metallicity relation of the local spiral galaxies.}
\end{figure}

We obtained an oxygen abundance of 12 + $\log$(O/H) = 8.98 $\pm$ 0.07\footnote[3]{All the metallicity values we use in this section for the DLA galaxy and other galaxies, are based on the R$_{23}$ index.} for this galaxy, for which the absolute magnitude has been estimated to be M$_B$(AB) = -21.1 (Chen {\it et al.} 2005). Figure \ref{fig:f4} shows the luminosity-metallicity relation for a large sample of various galaxies at z $<$ 0.25 (Tremonti {\it et al.} 2004). It is clear that the DLA galaxy falls close to the luminosity-metallicity relation of local spiral galaxies (Kobulnicky \& Zaritsky 1999; Zaritsky {\it et al.} 1994).

Recently \cite{sava05} investigated the metallicity of 69 galaxies at 0.4 $<$ z $< 1$ from the Gemini Deep Deep Survey and the Canada-France Redshift Survey. Using the R$_{23}$ metallicity indicator, they found the mean metallicity to be 12 + $\log$(O/H) = 8.78 $\pm$ 0.17, 0.05 dex higher than the value reported by \cite{kk04}. These surveys included all types of galaxies and did not show any redshift evolution. Our result from the R$_{23}$ indicator that the DLA galaxy has 12 + $\log$(O/H) = 8.98 $\pm$ 0.07 suggests that the DLA is a metal-rich galaxy and its metallicity lies at the high end of the metallicity distribution in \cite{sava05}.

The SFRs in various types of galaxies have been derived and discussed by several authors. These data show an enormous range, from virtually zero in gas-poor elliptical, S0, and dwarf galaxies to 10-20 M$_{\sun}$ yr$^{-1}$ in giant Sc galaxies. Much larger SFRs, up to  1000 M$_{\sun}$ yr$^{-1}$, can be found in starburst galaxies. The highest SFRs are associated with strong tidal interactions and mergers. SFRs of local blue compact galaxies, estimated by \cite{petr97}, are about 0.3-0.5 M$_{\sun}$ yr$^{-1}$. For LSB galaxies, \cite{vand00} estimated SFRs of about 0.03-0.2 M$_{\sun}$ yr$^{-1}$. For dwarf and LSB galaxies, \cite{vanz01} estimated SFRs from 10$^{-5}$ to 10$^{-1}$ M$_{\sun}$ yr$^{-1}$.

\begin{figure}
\centering
\includegraphics[scale=0.6]{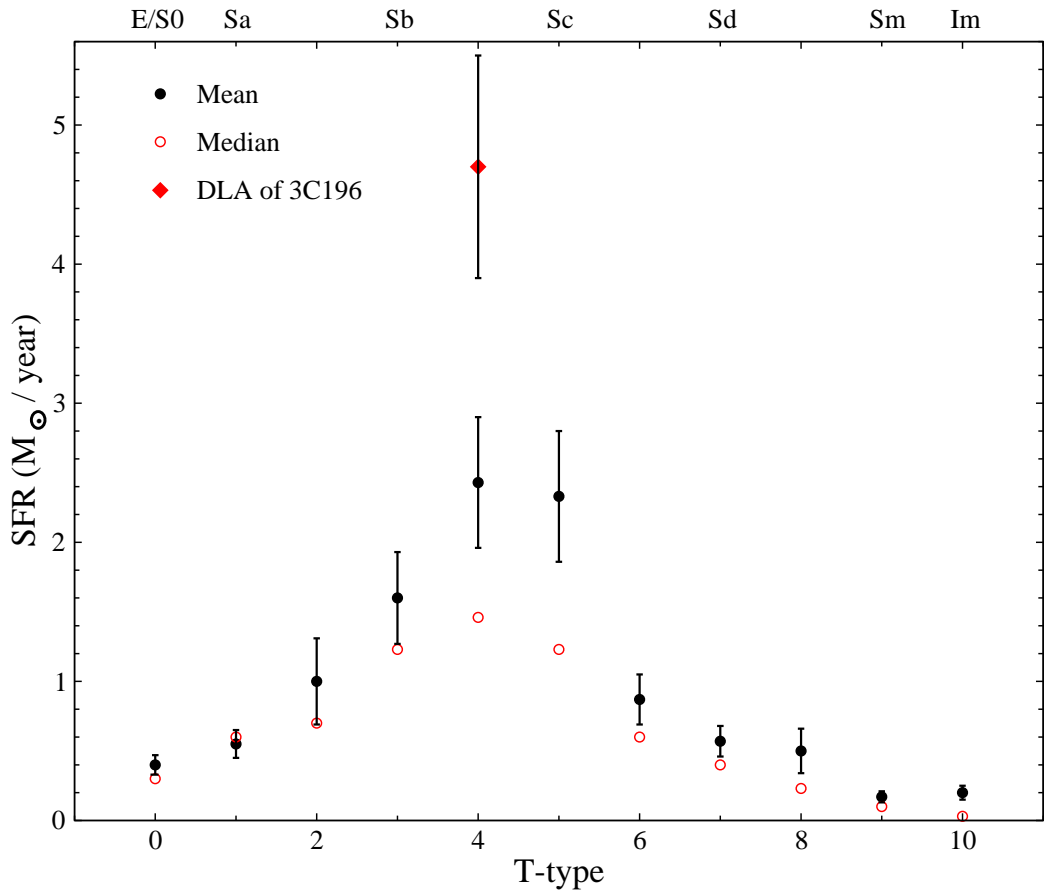}
\caption{\label{fig:f5} Mean and median star formation rates of 334 galaxies as a function of Hubble T-type from the compilation by \cite{jame04}. The error bars show the standard deviation.}
\end{figure}

In a study of star formation vs. galaxy morphology for 334 spiral and irregular nearby galaxies, \cite{jame04} found that the most strongly star-forming normal galaxies are those of Hubble types Sbc and Sc, with a wide range in SFR, up to 15 M$_{\sun}$ yr$^{-1}$. For local Sbc galaxies, the derived mean SFR is 2.4 $\pm$ 0.4 M$_{\sun}$ yr$^{-1}$ (see Figure \ref{fig:f5}) which is close to the mean calculated by \cite{kenn83} 3.1 $\pm$ 0.4 M$_{\sun}$ yr$^{-1}$ for this type\footnote[4]{Converted to the adopted Hubble constant of H$_o$ = 75 km s$^{-1}$ Mpc$^{-1}$ by \cite{jame04}.}. Although many of the galaxies in their survey appeared to have high SFRs, the derived mean is rather low, since very low SFRs are common for all Hubble types.

The SFR distribution of a sample of 168 low-mass IR-selected star-forming galaxies out to z=1.5 was explored by \cite{baue05}. In the redshift bin 0.4 $<$ z $<$ 1.0, they estimated SFR from 0.04 to 10 M$_{\sun}$ yr$^{-1}$, based on the [O II] $\lambda$3727 line. Their result suggests that the mean and the maximum SFR in each redshift bin increases with redshift.

Our result for the SFR of the DLA galaxy ($\approx$ 4.7 $\pm$ 0.8 M$_{\sun}$ yr$^{-1}$ for the SMC type extinction correction and 5.4 $\pm$ 1.0 M$_{\sun}$ yr$^{-1}$ for the MW type) agrees well with the expected SFR values for local Sbc galaxies and IR-selected star-forming galaxies in the redshift bin 0.4 $<$ z $<$ 1.0. This DLA appears to be a large galaxy with high SFR that lies at the high end of SFR distributions of \cite{jame04} and \cite{baue05}.

\subsection{Comparison with other DLAs and DLA models}
Metallicity studies indicate that most DLAs have lower mean metallicity than samples of other types of galaxies (Savaglio {\it et al.} 2005), and that they show little if any chemical evolution (e.g., Kulkarni {\it et al.} 2005; P\'eroux {\it et al.} 2006). The N$_{\rm HI}$-weighted mean Zn metallicity in the redshift bin of 0.1 $\lesssim$ z $\lesssim$ 1.4 reported by Kulkarni {\it et al.} 2005 is -0.94 $\pm$ 0.16. This would imply 12 + $\log$(O/H) = 7.75 $\pm$ 0.16, if we assume the Zn/O in DLAs is solar. Our DLA galaxy with 12 + $\log$(O/H) = 8.98$\pm$ 0.07 is much more metal-rich than the N$_{\rm HI}$-weighted mean.

\begin{figure}
\centering
\includegraphics[scale=0.4]{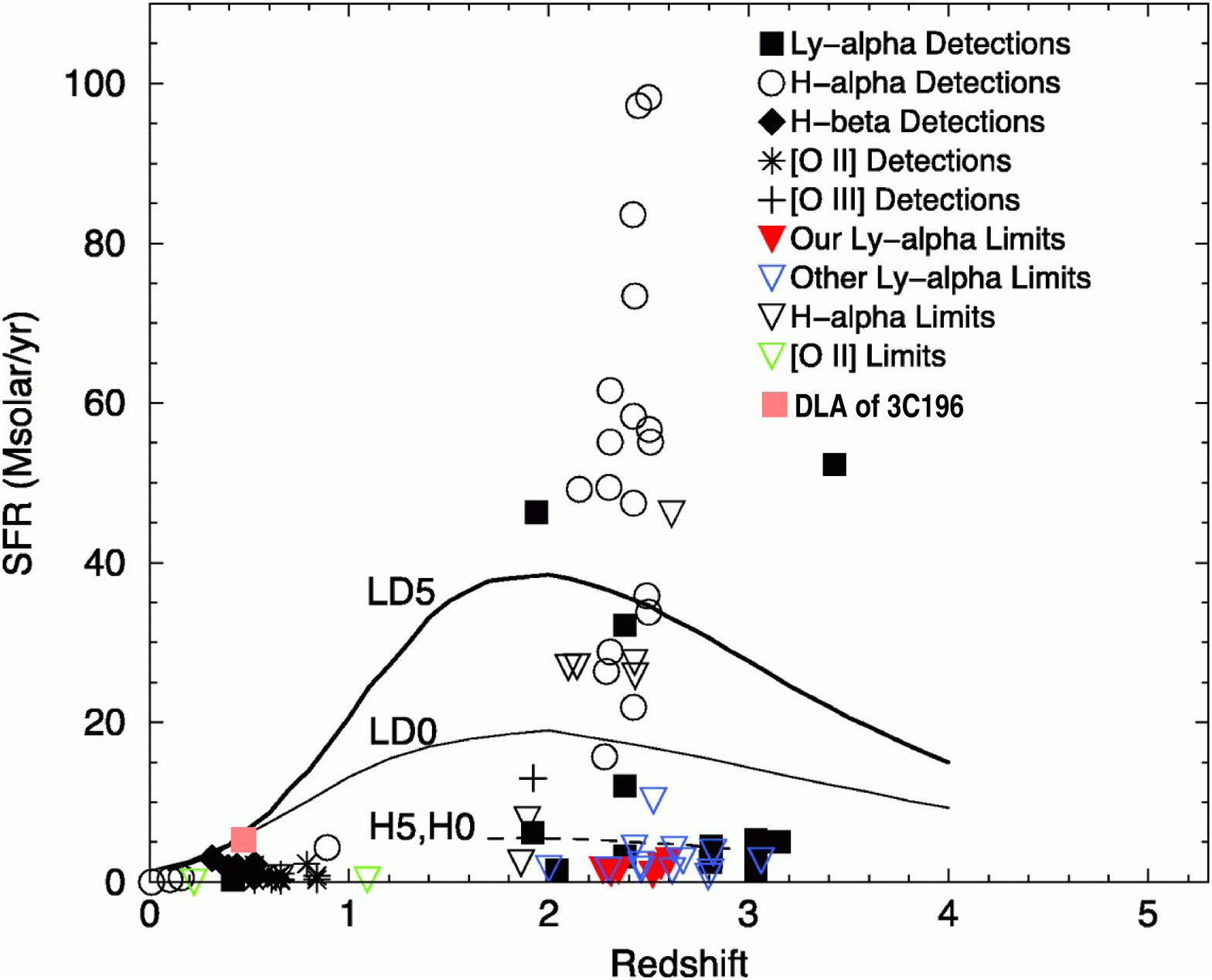}
\caption{\label{fig:f6} Star formation rate vs. redshift for candidate objects in QSO absorber fields from emission line imaging and spectroscopic searches for Ly$\alpha$, H$\alpha$, H$\beta$, [O II], and [O III], from \cite{kulk06} and references therein. The solid curves show the LD5 and LD0 ``large disk'' calculations of \cite{bunk99} (for q$_0$ = 0.5 and 0, respectively). The curve for $\Omega_{\Lambda}$ = 0.7, $\Omega_m$ = 0.3 should be between these two curves. The dashed curve shows their H5 and H0 predictions for the ``hierarchical'' hypothesis.}
\end{figure}

The star formation rates of DLAs are not well-understood. Emission line imaging searches for Ly$\alpha$, H$\alpha$, H$\beta$, [O II], and [O III] suggest low SFRs in a large fraction of absorption-selected galaxies (e.g., Kulkarni {\it et al.} 2006; see Figure \ref{fig:f6}). \cite{weat05} estimated SFRs from 9.5 to 28 M$_{\sun}$ yr$^{-1}$ from [O III] $\lambda$5007 emission in three DLAs at redshifts 1.9 $<$ z $<$ 3.2. Also, Wolfe {\it et al.} (2003a, 2003b) using the C II* absorption line method, estimated the SFRs per unit area of about 30 DLAs at z $\gtrsim$ 2 to be in the range 10$^{-3}$ to 10$^{-1}$ M$_{\sun}$ yr$^{-1}$ kpc$^{-2}$. These authors suggested that at high-redshift, DLAs are similar to the population of Lyman break galaxies. On the other hand, they noted that such large SFRs are inconsistent with the low metallicities seen in high-redshift DLAs. Indeed, the results of the C II* method are somewhat model dependent.

In fact, \cite{hopk05}, using N$_{\rm HI}$ to estimate the SFR suggested that DLAs have a dominant role in the SFR density only in the late universe (z $<$ 0.6), while at higher redshift their contribution is much less important. They suggested that the DLA population may be dominated by dwarf-like systems with low average SFRs or a late onset of star formation; in other words the DLAs identified so far may not account for the majority of the neutral gas at high redshift. However the assumption of \cite{hopk05} that the local SFR vs. surface density relation holds for DLAs has not been verified.

Recently, \cite{okos05} have investigated models for DLA absorbing galaxies at z $<$ 1, using a semi-analytic galaxy formation model. They predicted a broad SFR range spanning from 10$^{-6}$ to 10$^2$ M$_{\sun}$ yr$^{-1}$ and a mean value of 10$^{-2}$ M$_{\sun}$ yr$^{-1}$ for DLAs with z $<$ 1. Their study predicted that DLA galaxies brighter than L/L* $\sim$ 0.5 may have SFRs larger than 10 M$_{\sun}$ yr$^{-1}$. They concluded that most of DLAs at z $<$ 1 are LSB dwarf galaxies with low SFRs, although some are massive spiral galaxies.

Our SFR estimation of 4.7 M$_{\sun}$ yr$^{-1}$ for the DLA galaxy toward 3C196 lies within the broad SFR range predicted by \cite{okos05} model for DLA absorbing galaxies at z $<$ 1. However, the value is higher than that for a large fraction of DLAs: 63\% of the detections and about 73\% of the limits in the SFR sample studied by \cite{kulk06} have SFRs below 5 M$_{\sun}$ yr$^{-1}$. It is interesting to note that our value for the disk galaxy toward 3C196 agrees well with the ``large disk'' calculations of \cite{bunk99} (see Figure \ref{fig:f6}).

\section{Summary}

We have obtained spectra spanning 3500 \AA \ to 5700 \AA \ and 6700 \AA \ to 9700 \AA \ of a candidate DLA galaxy toward QSO 3C196. From emission line diagnostics, we found that the galaxy has a redshift of z = 0.4376 $\pm$ 0.0006. The DLA is bright (M$_B$(AB) = -21.1), metal rich (0.8 - 2 times solar metallicity) and has a SMC extinction-corrected SFR $\approx$ 4.7 $\pm$ 0.8 M$_{\sun}$ yr$^{-1}$ (MW extinction-corrected SFR $\approx$ 5.4 $\pm$ 1.0 M$_{\sun}$ yr$^{-1}$). This DLA thus appears to be a giant Sbc galaxy undergoing large SFR. This is the first case of a large, metal-rich galaxy apparently giving rise to a DLA and seems to be atypical as such a source. Finally, we can not rule out that galaxy \# 3, a dwarf galaxy which is slightly closer to the QSO and less luminous (M$_B$ = -22.0, Le Brun {\it et al.} 1997) than galaxy \# 4, is the DLA.

\acknowledgments
The authors wish to recognize and acknowledge the very significant cultural role and reverence that the summit of Mauna Kea has always had within the indigenous Hawaiian community. We are most fortunate to have the opportunity to conduct observations from this mountain. We also thank an anonymous referee for insightful comments that have improved the presentation of our work. SG and VPK gratefully acknowledge partial support from the National Science Foundation grants AST/0206197 and AST/0607739 (PI Kulkarni). M.T. acknowledges support from National Science Foundation grant AST-0205960.

\subsection*{APPENDIX A. \ Background Subtraction}
\begin{figure}
\begin{tabular}{cc}
\includegraphics[scale=0.5]{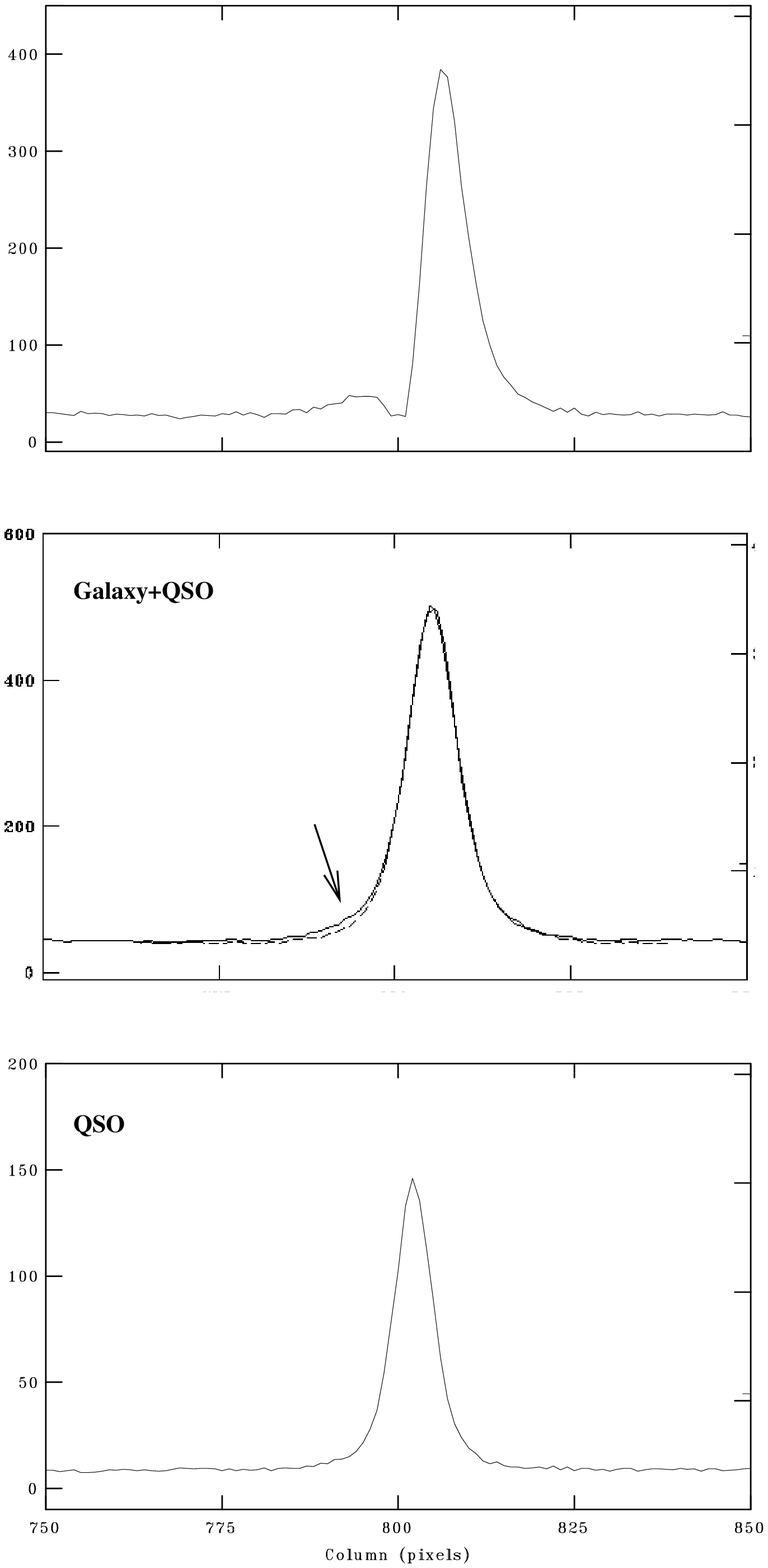}
\includegraphics[scale=0.5]{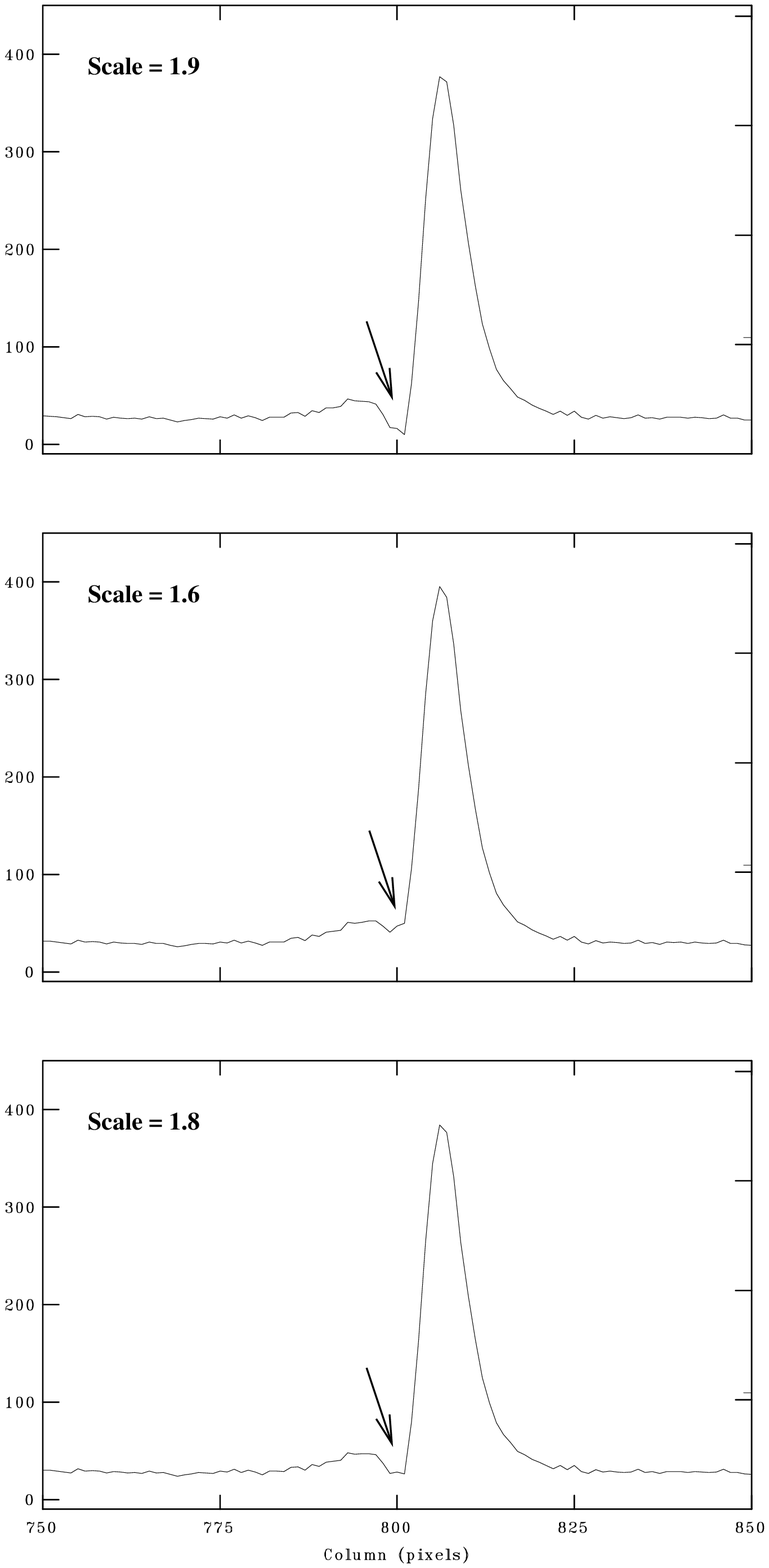}\\
\end{tabular}
\caption{\label{fig:f7}Left panels: The spatial profile along the dispersion axis in the blue channel. The lower panel shows the QSO profile and the middle Galaxy+QSO profile. The arrow points at the location of the asymmetry. The dashed line is a mirror image of the profile which enhances the asymmetry. We attribute the slight asymmetry of the Galaxy+QSO profile to the emission from the galaxy. The top panel shows the result of the subtraction of the shifted and scaled QSO template from the Galaxy+QSO.
Right panels: The spatial profile of the subtracted spectrum with 3 different scale factors. The bottom panel shows the best result in which the count level at the position of the gap (shown with arrow) matches the level of the overall background (scale = 1.8). In the middle panel, the scale factor is smaller than the best scale factor and the count level in the gap is above the level of the background (scale = 1.6). As a result the extracted spectrum is dominated by the QSO. This situation is reversed in the top panel, where the QSO is oversubtracted (scale = 1.9).}
\end{figure}

{\label{backsub}}
Here we describe the method developed to subtract the background contamination of the galaxy from the bright wings of the PSF of the QSO. The lower left panels of Figure \ref{fig:f7} show spatial profiles across the slit of the QSO and Galaxy+QSO respectively. A slight asymmetry in one of the sides of the Galaxy+QSO profile can be seen and we attribute this to the emission from the galaxy.

The goal of this method is to determine two parameters: 1) The amount of shift in the direction perpendicular to the dispersion axis of the QSO spectrum, to align the QSO spectrum with the Galaxy+QSO spectrum; 2) a factor to scale the flux of the QSO. To find out the correct values of the shift and the scale, we performed an iterative process. We started with a center to center alignment of the two profiles. The QSO spectrum was then subtracted from the Galaxy+QSO after rescaling the QSO template with various scale factors. The subtracted spectrum was extracted and examined. The result showed that center to center alignment of the profiles did not work, since large scale factors would remove flux from the galaxy, and small scale factors would not remove the QSO contamination from the galaxy spectrum.

We therefore allowed for spatial shifts and incorporated offset of $\sim$ 1 pixel between the centers of the two profiles. Again the QSO template was allowed to vary with different scale factors. With every subtraction, we examined the emission lines of extracted spectra as a figure of merit.

We performed this iteration process on each of the three exposures of 1800 s separately, since each individual exposure showed a profile with slightly different centers. The mean of the shifts was 0.4$''$. After subtracting the shifted QSO spectrum from the Galaxy+QSO, the galaxy and QSO emissions appeared distinct, with a minimum (i.e., a ``gap'') separating the two. The scale factor was set so that the count level at the position of the gap coincided with that of the background just outside the galaxy and QSO emissions. Larger scale factors than the one chosen would remove emission from the galaxy, while smaller ones would leave the QSO flux dominating the galaxy spectrum. The scale factor was different for each exposure with a mean of $\sim$ 2. 3 cases near the optimum value of the scale factor are shown in Figure \ref{fig:f7}.\\

\clearpage



\clearpage

\clearpage

\clearpage

\clearpage

\clearpage

\clearpage

\clearpage







\clearpage

\begin{deluxetable}{lcccccccc}
\tablecolumns{9}
\centering
\tablecaption{Identified emission lines in the DLA galaxy at z = 0.4376 \label{tab:res}}
\tablewidth{0pt}
\tablehead{
 \colhead{Line} & 
 \colhead{$\lambda_{rest}$} &
 \colhead{$\lambda_{obs}$} &
 \colhead{z} &
 \colhead{EW$_{obs}$} &
 \colhead{Flux} &
 \colhead{L} &
 \multicolumn{2}{c}{L$^o$}\\
\cline{8-9}\\
\colhead{} &
 \colhead{} &
 \colhead{} &
 \colhead{} &
 \colhead{} &
 \colhead{} &
 \colhead{} &
 \colhead{SMC} &
 \colhead{MW}\\
\cline{1-9}\\
\colhead{} &
 \colhead{} &
 \colhead{} &
 \colhead{} &
 \colhead{} &
 \colhead{10$^{-17}$} &
 \colhead{10$^{40}$} &
 \multicolumn{2}{c}{10$^{40}$}\\
\colhead{} &
 \colhead{(\AA )} &
 \colhead{(\AA )} &
 \colhead{} &
 \colhead{(\AA )} &
 \colhead{(ergs/s/cm$^{2}$)} &
 \colhead{(ergs/s)} &
 \multicolumn{2}{c}{(ergs/s)}}
\startdata
[O II] &3727&5360.06&0.43830&6.6 $\pm$ 0.6&4.6 $\pm$ 0.4&3.2 $\pm$ 0.3&31.2 $\pm$ 10.5&34.0 $\pm$ 11.9\\

H $\beta$ &4861&6992.69&0.43843&6.7 $\pm$ 0.4&5.6 $\pm$ 0.3&3.9 $\pm$ 0.2&20.5$\pm$ 5.0&23.5 $\pm$ 6.2\\

[O III]\tablenotemark{\dag} &4959&7118.15&0.43540&0.9 $\pm$ 0.3&0.8 $\pm$ 0.2&0.5 $\pm$ 0.2&2.6 $\pm$ 1.1&3.0 $\pm$ 1.3\\

[O III]\tablenotemark{\dag} &5007&7203.61&0.43875&1.7 $\pm$ 0.3&1.5 $\pm$ 0.2&1.0 $\pm$ 0.2&5.1 $\pm$ 1.4&5.8 $\pm$ 1.7\\

[He I]\tablenotemark{\dag} &5876&8446.88&0.43752&1.9 $\pm$ 0.6&1.6 $\pm$ 0.5&1.1 $\pm$ 0.4&4.0 $\pm$ 1.5&4.6 $\pm$ 1.8\\

[O I]\tablenotemark{\dag} &6300&9045.69&0.43576&1.4 $\pm$ 0.3&1.2 $\pm$ 0.3&0.9 $\pm$ 0.2&2.7 $\pm$ 0.8&3.1 $\pm$ 1.0\\

H $\alpha$ &6563&9433.65&0.43744&26.2 $\pm$ 1.5&28.3 $\pm$ 1.6&19.8 $\pm$ 1.1&68.2 $\pm$ 10.0&59.5 $\pm$ 12.7\\

[N II]\tablenotemark{\dag} &6584&9463.57&0.43749&7.1 $\pm$ 1.5&7.6 $\pm$ 1.6&5.4 $\pm$ 1.1&16.0 $\pm$ 4.2&18.3 $\pm$ 5.1\\
\enddata
\tablenotetext{\dag}{These features are suggestive but should be regarded as tentative, because they are comparable to the noise level.}

\end{deluxetable}
\clearpage

\begin{deluxetable}{ccccccccccccc}
\tablecolumns{13}
\centering
\tablecaption{Summary of properties of the DLA galaxy \label{tab:sum}}
\tablewidth{0pt}
\tablehead{
 \colhead{z$_{abs}$} & 
 \colhead{z$_{em}$} &
 \colhead{M$_B$} &
 \colhead{} &
 \multicolumn{2}{c}{E(B$-$V)} &
 \colhead{} &
 \multicolumn{3}{c}{12 + $\log$(O/H)} &
 \colhead{} &
 \multicolumn{2}{c}{SFR}\\
\colhead{} &
 \colhead{} &
 \colhead{} &
 \colhead{} &
 \colhead{} &
 \colhead{} &
 \colhead{} &
 \colhead{} &
 \colhead{} &
 \colhead{} &
 \colhead{} &
 \multicolumn{2}{c}{(M$_{\sun}$ yr$^{-1}$)}\\
\cline{5-6} \cline{8-10} \cline{12-13}\\
\colhead{} &
 \colhead{} &
 \colhead{} &
 \colhead{} &
 \colhead{SMC} &
 \colhead{MW} &
 \colhead{} &
 \colhead{R$_{23}$} &
 \colhead{N2} &
 \colhead{O3N2} &
 \colhead{} &
 \colhead{SMC} &
 \colhead{MW}}
\startdata
0.4371 &0.4376 &-21.1 &&0.54 &0.56 &&8.98 &8.56 &8.74 &&4.7 &5.4\\
$\pm$ 0.0003 &$\pm$ 0.0006 &&&$\pm$ 0.08 &$\pm$ 0.08 &&$\pm$ 0.07 &$\pm$ 0.12 &$\pm$ 0.07 &&$\pm$ 0.8 &$\pm$ 1.0
\enddata
\end{deluxetable}
\clearpage


\begin{thebibliography}{}

\bibitem[Allende-Prieto {\it et al.}(2001)]{alen01} Allende-Prieto, C., Lambert, D. L., \& Asplund, M. 2001, ApJ, 556, L63

\bibitem[Asplund {\it et al.}(2004)]{aspl04} Asplund, M., Grevesse, N., Sauval, A. J., Alende-Prieto, C., \& Kiselman, D. 2004, A\&A, 417, 751

\bibitem[Bauer {\it et al.}(2005)]{baue05} Bauer, A. E., Drory, N., Hill, G. J., \& Feulner, G. 2005, ApJ, 621, L89

\bibitem[Boiss\'e {\it et al.}(1998)]{bois98} Boiss\'e, P., Le Brun, V., Bergeron, J., \& Deharveng, J. M. 1998, A\&A, 333, 841

\bibitem[Boiss\'e \& Boulade(1990)]{bois90} Boiss\'e P., \& Boulade O. 1990, A\&A, 236, 291

\bibitem[Brown \& Mitchell(1983)]{brmi83} Brown, R. L., \& Mitchell, K. J. 1983, \apj, 264, 87

\bibitem[Bunker {\it et al.}(1999)]{bunk99} Bunker, A. J., Warren, S. J., Clements, D. L., Williger, G. M., \& Hewett, P. C. 1999, MNRAS, 309, 875


\bibitem[Cardelli {\it et al.}(1989)]{card89} Cardelli, J. A., Clayton, G. C., \& Mathis, J. S. 1989, ApJ, 345, 245

\bibitem[Chen {\it et al.}(2005)]{chen05} Chen, H.-W., Kennicutt, R. C., Jr., \& Rauch, M. 2005, ApJ, 620, 703

\bibitem[Chun {\it et al.}(2006)]{chun06} Chun, M., Gharanfoli, S., Kulkarni, V. P., \& Takamiya, M. 2006, AJ in press, Vol 131

\bibitem[Cohen {\it et al.}(1996)]{cohe96} Cohen, R. D., Beaver, E. A., Diplas, A., Junkkarinen, V. T., Barlow, T. A., \& Lyons, R. W. 1996, ApJ, 456, 132

\bibitem[Fall \& Pei(1993)]{fape93} Fall, S. M., \& Pei, Y. C. 1993, ApJ, 402, 479

\bibitem[Fukugita {\it et al.}(1995)]{fuku95} Fukugita, M., Shimasaku, K., \& Ichikawa, T. 1995, PASP, 107, 945

\bibitem[Hopkins {\it et al.}(2005)]{hopk05} Hopkins, A. M., Rao, S. M., \& Turnshek D. A. 2005, ApJ, 630, 108

\bibitem[James {\it et al.}(2004)]{jame04} James, P. A., {\it et al.} 2004, A\&A, 414, 23

\bibitem[Jimenez {\it et al.}(1999)]{jime99} Jimenez, R., Bowen, D. V., \& Matteucci, F. 1999, ApJ, 514, L83

\bibitem[Kennicutt(1998)]{kenn98} Kennicutt, R. C., Jr. 1998, ARA\&A, 36, 189

\bibitem[Kennicutt(1983)]{kenn83} Kennicutt, R. C., Jr. 1983, ApJ, 272, 54

\bibitem[Kobulnicky \& Kewley(2004)]{kk04} Kobulnicky, H. A., \& Kewley, L. J. 2004, ApJ, 617, 240

\bibitem[Kobulnicky {\it et al.}(1999)]{kobu99} Kobulnicky, H. A., Kennicutt, R. C., Jr., \& Pizagno, J. L. 1999, ApJ, 514, 544

\bibitem[Kobulnicky \& Zaritsky(1999)]{koza99} Kobulnicky, H. A., \& Zaritsky, D. 1999, ApJ, 511, 118

\bibitem[Kormendy \& Kennicutt(2004)]{koke04} Kormendy, J., \& Kennicutt, R. C., Jr. 2004, ARA\&A, 42, 603

\bibitem[Kulkarni {\it et al.}(2006)]{kulk06} Kulkarni, V. P., Woodgate, B. E., York, D. G., Thatte, D. G., Meiring, J., Palunas, P., \& Wassell, E. 2006, ApJ, 636, 30

\bibitem[Kulkarni {\it et al.}(2005)]{kulk05} Kulkarni, V. P., Fall, S. M., Lauroesch, J. T., Khare, P., \& Truran, J. W. 2005, ApJ, 618, 68

\bibitem[Kulkarni {\it et al.}(2001)]{kulk01} Kulkarni, V. P., Hill, J. M., Schneider, G., Weymann, R. J., Storrie-Lombardi, L. J., Rieke, M. J., Thompson, R. I., \& Jannuzi, B. 2001, ApJ, 551, 37

\bibitem[Kulkarni {\it et al.}(2000)]{kulk00} Kulkarni, V. P., Hill, J. M., Schneider, G., Weymann, R. J., Storrie-Lombardi, L. J., Rieke, M. J., Thompson, R. I., \& Jannuzi, B. 2000, ApJ, 536, 36

\bibitem[Le Brun {\it et al.}(1997)]{lebr97} Le Brun, V., Bergeron, J., Boiss\'e, P., \& Deharveng, J. M. 1997, A\&A, 321, 733

\bibitem[Lytle {\it et al.}(1999)]{lytl99} Lytle, D., Stobie, E., Ferro, A., \& Barg, I. 1999, ASPC, 172, 445

\bibitem[McGaugh(1991)]{mcga91} McGaugh, S. S. 1991, ApJ, 380, 140

\bibitem[Nagamine {\it et al.}(2004)]{naga04} Nagamine, K., Springel, V., \& Hernquist, L. 2004, MNRAS, 348, 435

\bibitem[Oke {\it et al.}(1995)]{oke95} Oke, J. B., {\it et al.} 1995, PASP, 107, 375

\bibitem[Oke(1990)]{oke90} Oke, J. B. 1990, AJ, 99, 1621

\bibitem[Oke \& Gunn(1983)]{oke83} Oke, J. B., \& Gunn, J. E. 1983, ApJ, 266, 713

\bibitem[Okoshi \& Nagashima(2005)]{okos05} Okoshi, K., \& Nagashima, M. 2005, ApJ, 623, 99

\bibitem[Pagel {\it et al.}(1979)]{page79} Pagel, B. E. J., Edmunds, M. G., Blackwell, D. E., Chun, M. S., \& Smith, G. 1979, MNRAS, 189, 95

\bibitem[Pei(1992)]{pei92} Pei, Y. C. 1992, \apj, 395, 130

\bibitem[P\'eroux {\it et al.}(2006)]{pero06} P\'eroux, C., Kulkarni, V. P., Meiring, J., Ferlet, R., Khare, P., Lauroesch, J. T., Vladilo, G., \& York, D. G. 2006, MNRAS, submitted

\bibitem[Petrosian {\it et al.}(1997)]{petr97} Petrosian, A. R., Boulesteix, J., Comte, G., Kunth, D., \& Leoarer, E. 1997, A\&A, 318, 390

\bibitem[Pettini \& Pagel(2004)]{pepa04} Pettini, M., \& Pagel, B. E. J. 2004, MNRAS, 348, L59

\bibitem[Pettini {\it et al.}(1997)]{pett97} Pettini, M., Smith, L. J., King, D. L., \& Hunstead, R. W. 1997, \apj, 486, 665

\bibitem[Prochaska \& Wolfe(2002)]{prwo02} Prochaska, J. X. \& Wolfe, A. M. 2002, \apj, 566, 68

\bibitem[Richards {\it et al.}(2003)]{rich03} Richards, G. T., {\it et al.} 2003, AJ, 126, 1131

\bibitem[Richards {\it et al.}(2001)]{rich01} Richards, G. T., {\it et al.} 2001, AJ, 121, 2308

\bibitem[Savaglio {\it et al.}(2005)]{sava05} Savaglio, S., {\it et al.} 2005, ApJ, 635, 260


\bibitem[Tremonti {\it et al.}(2004)]{trem04} Tremonti, C. A., {\it et al.} 2004, \apj, 613, 898

\bibitem[van den Hoek {\it et al.}(2000)]{vand00} van den Hoek, L. B., de Blok, W. J. G., van der Hulst, J. M., \& de Jong, T. 2000, A\&A, 357, 397

\bibitem[van Zee (2001)]{vanz01} van Zee, L. 2001, AJ, 121, 2003

\bibitem[Weatherley {\it et al.}(2005)]{weat05} Weatherley, S. J., Warren, S. J., Møller, P., Fall, S. M., Fynbo, J. U., \& Croom, S. M. 2005, MNRAS, 358, 985

\bibitem[Wolfe {\it et al.}(2003a)]{wolf03a} Wolfe, A. M., Prochaska, J. X., \& Gawiser E. 2003a, ApJ, 593, 215 

\bibitem[Wolfe {\it et al.}(2003b)]{wolf03b} Wolfe, A. M., Prochaska, J. X., \& Gawiser E. 2003b, ApJ, 593, 235

\bibitem[York {\it et al.}(2006)]{york06} York, D. G., {\it et al.} 2006, MNRAS, 367, 945

\bibitem[Zaritsky {\it et al.}(1994)]{zari94} Zaritsky, D., Kennicutt, R. C., Jr., \& Huchra, J. P. 1994, ApJ, 420, 87

\end{thebibliography}
\end{document}